\documentclass[prb,twocolumn,showpacs,preprintnumbers,amsmath,amssymb]{revtex4}

\usepackage{graphicx}
\usepackage{dcolumn}
\usepackage{bm}
\def\S{{\bf S}}

\begin{document}
\title{Two quantum spin models on the checkerboard lattice with an exact two-fold
degenerate Shastry-Sutherland ground state}
\author{Brijesh Kumar}
 \email{brijesh.kumar@epfl.ch}
 \affiliation{Institute of Theoretical Physics, \'Ecole Polytechnique F\'ed\'erale de Lausanne, CH-1015 
              Lausanne, Switzerland}
 \date{\today} 
\begin{abstract}
Two quantum spin models with bilinear-biquadratic exchange interactions are constructed on the 
checkerboard lattice. It is proved that, under certain sufficient conditions on the exchange 
parameters, their ground states consist of two degenerate Shastry-Sutherland singlet 
configurations. The constructions are studied for arbitrary spin-S. The sufficient conditions 
for the existence of ferromagnetic ground state are also found exactly. The approximate quantum 
phase diagrams are presented using the exact results, together with a variational estimate for 
the N\'eel antiferromagnetic phase. A two-leg spin-1/2 ladder model, based on one of the above 
constructions, is considered which admits exact solution for a large number of eigenstates. 
The ladder model is shown to have exact level-crossing between the rung-singlet state and the 
AKLT state in the singlet ground state. Also introduced is the notion of perpendicularity for  
quantum spin vectors, which appears in the discussion on one of the two checkerboard models,
and is discussed in the Appendix.
\end{abstract}

\maketitle

\section{\label{sec:intro} Introduction\protect}
The area of frustrated quantum magnetism is of great current interest. The increasing
number of real materials with frustrated spin interactions requires us to have 
better insight into the nature of possible spin quantum states that may arise due to frustration, 
and the excitations thereof. While the phenomenological studies are guided by experimental findings, 
there is also a formal interest in the subject with an aspiration for constructing exactly
solvable models, and for investigating possibly realizable new physics. The Shastry-Sutherland
(SS) model~\cite{SS} of frustrated quantum spins, with a surprising realization in 
SrCu$_2$(BO$_3$)$_2$~\cite{kageyama,miyahara_ueda}, is an interesting example of formal 
studies making sense in real systems.

\begin{figure}
\includegraphics[width=5cm]{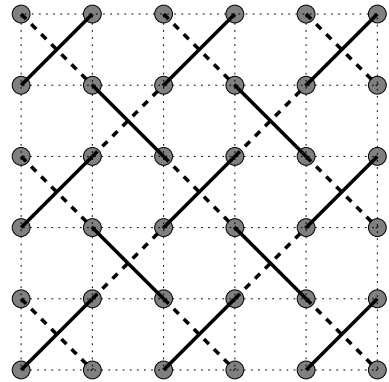}
\caption{\label{fig:2SS} Two SS singlet states on a checkerboard. A solid (or dashed) line
joining two diagonally opposite sites denotes the singlet state. The solid dimer-singlets 
form one of the SS states, and the dashed dimers form the other. Thin dotted lines just mark 
the underlying square lattice.}
\end{figure}

The SS model is described by an antiferromagnetic bilinear spin exchange Hamiltonian on square 
lattice with the nearest neighbor ({\em nn}) interaction, and with the next nearest neighbor 
interaction along a select choice of diagonal bonds (see the arrangement of either 
solid or dashed diagonal bonds in Fig.~\ref{fig:2SS}). The lattice with such a specific topology 
of connections is called the SS lattice. When the diagonal exchange is twice as big or bigger 
than the {\em nn} exchange, then the configuration consisting of dimer-singlets on the diagonal 
bonds of the SS lattice is an exact ground state of the model. We call this ground state as the
SS singlet state (or simply the SS state). The SS state, however, continues to be the ground 
state for lower values of the 
diagonal exchange, approximately upto 1.48 times the {\em nn} exchange, below which the system
undergoes a quantum phase transition into a new state (and finally into the N\'eel
antiferromagnetic state), as suggested by various numerical studies~\cite{mila,koga_kawakami}.

In recent times, there has also been a renewed interest in the studies of spin models with 
multiple exchange interactions (that is, more than bilinear exchange), particularly due to some 
suggestions that such higher order exchange interactions may be relevant in the insulating,
antiferromagnetic phase of the high-T$_C$ cuprates~\cite{coldea}, and also in the spin 
ladders~\cite{spin_ladder}. However, the models with multiple exchange have been studied for 
various independent reasons. Historically, the physical importance of multiple exchange was first 
realized in the studies of magnetism in solid $^3$He, starting effectively with the work of 
Thouless~\cite{thouless,He3}. Even for strongly correlated electronic systems, from the point of 
view of the Hubbard model, the presence of multiple exchange spin interactions is rather 
apparent from the fourth order perturbation theory~\cite{takahashi,macdonald} in $t/U$. The 
actual physical relevance of such higher order exchange interactions, however, may differ in 
different materials. 

In spin systems, a rather well known example of the models with higher order exchange is the 
Affleck, Kennedy, Lieb and Tasaki (AKLT) construction~\cite{AKLT_Lett, AKLT}. These are exactly 
solvable models with valence-bond-solid ground states. In one dimension, the AKLT model is of 
particular interest, as it explicitly demonstrates the Haldane spin-gap 
conjecture~\cite{Haldane1,Haldane2} for a spin-1 chain. There have also been studies of the 
models with multiple exchange interactions on the plaquettes of a two-leg spin 
ladder~\cite{andreas}, some of which allow exact solution for the ground 
state~\cite{Mikeska1, Mikeska2} (notably that of the AKLT type, besides the rung-singlet and 
the ferromagnetic ground states).

In this paper, we construct two quantum spin models with bilinear and biquadratic exchange 
interactions on the checkerboard lattice, and show that they have an exact ground state 
consisting of two SS singlet states. In Section~\ref{sec:models}, we motivate and define these 
models. In Sections~\ref{sec:model1} and \ref{sec:model2}, we describe the proofs of their 
exact ground states, and also discuss their quantum phase diagrams. Inspired by our discussion 
on one of the two models, we introduce the notion of perpendicularity for quantum spin vectors, 
which is discussed in the Appendix. We also study a two-leg spin-1/2 ladder model in 
Section~\ref{sec:ladder}, which incidentally allows the exact solution of a large number of 
eigenstates. Interestingly, the AKLT singlet state (as for a spin-1 chain) becomes an exact 
ground state of the ladder model which shows level-crossing transition between the rung-singlet 
and the AKLT states in the singlet ground state. The ladder model is shown to have another ground 
state with alternating singlet-triplet arrangement on rungs, which gives rise to extensive entropy 
in this ground state. Finally, we conclude with a summary, and with some general remarks.
\section{\label{sec:models} The Models}
\begin{figure}
\includegraphics[width=5cm]{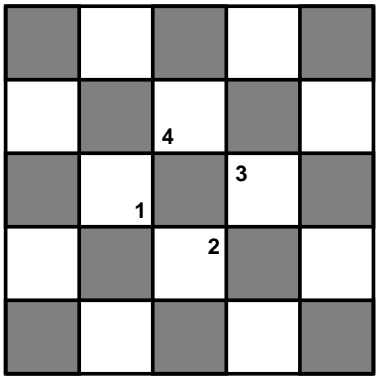}
\caption{\label{fig:lattice} The checkerboard lattice. The models considered in the text
consist of four-spin block-Hamiltonians sitting on the dark plaquettes of the lattice. Also
shown is the convention of spin labeling on a single plaquette.}
\end{figure}

The construction of models, that are presented in this paper, is inspired by a simple
observation that a SS state is also a zero energy eigenstate of the {\em nn} Heisenberg 
model on square lattice, but it is not the ground state. This fact is already contained in 
the proof that the SS state is always an eigenstate of the SS model~\cite{SS}. Since the 
{\em nn} Heisenberg model on square lattice is same as the SS model with zero diagonal 
exchange interaction, there are four degenerate SS eigenstates for the {\em nn} Heisenberg 
model, two for each of the two sets of checkerboard plaquettes (say, black or white 
plaquettes in Fig.~\ref{fig:lattice}).

Now the question is, `` how do we make these SS states also the ground state ? '' One simple 
answer is the SS model itself. But the SS lattice has lower translational symmetry as compared 
to the square lattice (or even the checkerboard lattice), and therefore it can have only one 
of the four SS states as the ground state. We wish to keep at least the checkerboard (if not 
the full square lattice) symmetry in our model, and therefore allow the possibility of two  
degenerate SS singlet configurations in the ground state (though the original motivation was 
to find a  model with four-fold degenerate SS ground state).  

For our discussion, we write the Hamiltonian of the $nn$ Heisenberg model as: 
H$=J\sum_R h(R,R+x,R+y,R+x+y)$, where the summation over $R$ is done for one sublattice only, 
and the block-Hamiltonian $h(R,R+x,R+y,R+x+y)$ acts on a plaquette of the corresponding 
checkerboard lattice. For the convenience of writing, on a given plaquette, we denote the 
site indices $R$, $R+x$, $R+x+y$ and $R+y$ as 1, 2, 3, and 4 respectively (see 
Fig.~\ref{fig:lattice}). In this notation, \(h(R,R+x,R+y,R+x+y) \equiv 
h = (\S_1\cdot\S_2 + \S_2\cdot\S_3 + \S_3\cdot\S_4 + \S_4\cdot\S_1)\). 
Denoting the total spin on the diagonal bonds of a plaquette as 
$\S_{13} = \S_1 + \S_3$ and $\S_{24} = \S_2 + \S_4$ gives $h=\S_{13}\cdot\S_{24}$. 

It is evident that $h$ annihilates any state with a singlet on either of the two diagonal 
bonds of the corresponding plaquette. This fact gives rise to four degenerate SS eigenstates 
with zero energy for the square lattice {\em nn} Heisenberg model. If we suitably modify our
block-Hamiltonian such that the zero energy states of $h$ become the lowest energy states of 
the new block-Hamiltonian, then the lower bound to the ground state energy of the 
corresponding Hamiltonian on the checkerboard will also be zero. This will give us a ground 
state with two degenerate SS states. We realize that there are at least two simple ways of 
achieving our purpose on the checkerboard lattice. These are described below.
\subsection{\label{model1} Definition of Model-\texttt{I}} 
One way of constructing a model with the ground state consisting of two degenerate SS states, 
is to introduce an additional term proportional to $(\S_{13}\cdot\S_{24})^2$ in the previously 
discussed block-Hamiltonian $h$, and put these new block-Hamiltonians on the plaquettes of a 
checkerboard to make a lattice Hamiltonian. To be explicit, the new block-Hamiltonian, 
$h_\texttt{I}$, is given as:
\begin{equation}
h_\texttt{I} = J~h + K~h^2~~~\mbox{where}~~h=\S_{13}\cdot\S_{24}
\label{eq:h1}
\end{equation}
Evidently, the eigenstates of $h_\texttt{I}$ are completely given by the eigenstates of $h$, and 
specified by the quantum numbers S$_{13}$ and S$_{24}$ of the spins on the diagonals and the 
total block-spin, S$_{tot}$. The corresponding Hamiltonian on the checkerboard lattice is given 
as : 
\begin{equation}
\mbox{H}_\texttt{I} = \sum_{R} h_\texttt{I}(R,R+x,R+x+y,R+y)
\label{eq:model1}
\end{equation}

Since the eigenvalues of $h^2$ will always be greater than or equal to zero, for sufficiently 
$+ve$ $K$, $h_\texttt{I}$ will like to have zero energy eigenstates of $h$ as the ground state 
(see Table~\ref{tab:h1}). This, for the lattice model H$_\texttt{I}$, will give us two-fold 
degenerate SS ground state. The details will be discussed in Section~\ref{sec:model1}.
\subsection{\label{model2} Definition of Model-\texttt{II}}
The second model is constructed using the following choice of the block-Hamiltonian.
\begin{equation}
h_\texttt{II} = J~h + \frac{K}{4}~\S_{13}^2~\S_{24}^2
\label{eq:h2}
\end{equation} 
Again, the block-states are specified by S$_{13}$, S$_{24}$ and S$_{tot}$.
In this case, the term proportional to $K$ is the interaction between the non-singlet states of 
the diagonal bonds of a plaquette. The lattice model is constructed by assigning an 
$h_\texttt{II}$ to each of the corner sharing plaquettes on a checkerboard. The Hamiltonian of 
our second model is given as: 
\begin{equation}
\mbox{H}_\texttt{II} = \sum_{R} h_\texttt{II}(R,R+x,R+x+y,R+y)
\label{eq:model2}
\end{equation}
(The summation over $R$ in Eqs.~\ref{eq:model1} and \ref{eq:model2} is done only on one 
sublattice of the underlying square lattice.)
As we will see in Section~\ref{sec:model2}, for sufficiently strongly repulsive interaction $K$, 
the two SS states form an exact ground state of H$_\texttt{II}$. 
\begin{table}
\caption{\label{tab:h1} Eigen-spectrum spectrum of $h_\texttt{I}$ for S=1/2.}
\begin{ruledtabular}
\begin{tabular}{ccl}
S$_{13}$ & S$_{24}$ & Energy eigenvalues\\ \hline
0 & 0 & ~0\\
0 & 1 & ~0\\
1 & 0 & ~0\\ \hline
1 & 1 & \begin{tabular}{rcl} $-2J + 4K$ & : & S$_{tot} = 0$\\
                           $-J + K$ & : & S$_{tot} = 1$\\
                           $J + K$ & : & S$_{tot} = 2$
        \end{tabular}
\end{tabular}
\end{ruledtabular}
\end{table}

\section{\label{sec:model1} Discussion on Model-\texttt{I}}
{\bf Ground State for Spin-1/2 Case :} In order to find the exact ground state of the 
Hamiltonian H$_\texttt{I}$, consider the eigen-spectrum of the block-Hamiltonian $h_\texttt{I}$. 
For spin-1/2, the eigenstates of $h_\texttt{I}$ are given in Table~\ref{tab:h1}. It is 
evident that the zero energy eigenstates of $h_\texttt{I}$ will become 
the lowest in energy for $K>|J|$. We get this condition on $J$ and $K$ by demanding that the 
eigenvalues in (S$_{13}$, S$_{24}$)=(1,1) sector are greater than zero.

Under this condition, each plaquette of the checkerboard can simultaneously attain its 
lowest energy ($i.e.$, zero) eigenstate by forming a singlet on at least one of the two 
diagonal bonds. We may call the singlet on a diagonal bond a {\em diagonal-singlet}. 
The prescription of having one diagonal-singlet per plaquette generates two degenerate 
SS singlet configurations on a checkerboard (see Fig.~\ref{fig:2SS}), which in the present 
case will form zero energy eigenstates of H$_\texttt{I}$. These two SS states will also 
form the ground state, because the lower bound on the eigenvalues of H$_\texttt{I}$ is zero 
for $K>|J|$. 

If we try to construct a singlet configuration by allowing some plaquettes to have two
diagonal-singlets, then it will leave some other plaquettes unsatisfied. Consider a 
plaquette on the checkerboard with singlets on both the diagonal bonds. Each of the four 
plaquettes that share a
corner with this plaquette can attain zero energy state only by having a singlet on the
diagonal bond that does not have the shared corner (spin). This implies that another
plaquette, which shares corners with two of these (four plaquettes), will not be able to 
form a singlet along any of its two diagonals. Thus, the two SS configurations are the 
only ones where each plaquette is simultaneously satisfied, and hence a two-fold 
degenerate ground state.

For $J<0$ and $J < K < -J$, H$_\texttt{I}$ has an exact ferromagnetic (FM) ground state, 
with $K=|J|$ as the line of level-crossing between the SS and the FM ground states. This 
condition for the existence of the FM ground state can again be inferred by demanding that 
the energy of S$_{tot}=2$ in (1,1) sector of Table~\ref{tab:h1} be smaller than rest of the 
energies. Again each plaquette on the checkerboard can attain its lowest energy state by 
polarizing all the spins in the same direction, thus giving rise to the FM ground state. 
Next we find the exact conditions for the SS and the FM ground states for spin-S system.
\begin{figure}
\includegraphics[width=5.5cm]{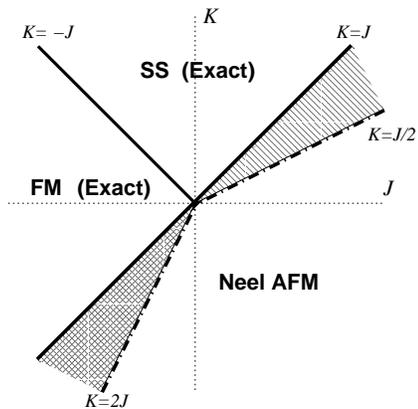}
\caption{\label{fig:phase_m1} The quantum phase diagram of model H$_\texttt{I}$ for S=1/2.
$K=-J$ is the level-crossing phase boundary between the exact SS and FM ground states. The 
SS ground state consists of two degenerate SS singlet states. The two lines, $K=J/2$ for 
$J>0$ and $K=2J$ for $J<0$, define the (variational) phase boundary for N\'eel 
antiferromagnetic (AFM) state. The nature of the ground state in the shaded regions is not 
known.}
\end{figure}

{\bf Ground State for Spin-S Case :} 
Let us denote an eigenvalue of $h$ as $\varepsilon$ and that of $h_\texttt{I}$ as 
$\varepsilon_\texttt{I}$. From the definition of $h_\texttt{I}$ (as given in Eq.~\ref{eq:h1}),
$\varepsilon_\texttt{I} = J~\varepsilon + K~\varepsilon^2$. And as usual, all the states of 
$h_\texttt{I}$ are labeled by spin quantum numbers S$_{13}$, S$_{24}$ and S$_{tot}$.
Let S$_{13}$, S$_{24}$ and S$_{tot}$ take the values $m$, $n$, and $l$ respectively, where $m$ 
and $n$ take integer values between 0 and 2S, and $l$ is an integer such that \(|m-n|\leq l
\leq m+n\). For a given choice of $l$, $m$ and $n$, 
$\varepsilon = \{l(l+1) - m(m+1) - n(n+1)\}/2$, which in turn gives us the value of 
$\varepsilon_\texttt{I}$. Notice that $\varepsilon$ will always be an integer (since all the 
terms inside $\{~\}$ are even integers).

Now we ask when the states with a diagonal-singlet ($m$ or $n=0$) become the ground 
state of $h_\texttt{I}$. For $m$ (or $n$) $=0$, $\varepsilon_\texttt{I}=0$ (since $l$ 
is same as $n$ (or $m$)). Therefore, we need to find the conditions on $J$ and $K$ such 
that $\varepsilon_\texttt{I}\geq 0$. There are two things to be done. Since any state with 
$\varepsilon =0$ gives $\varepsilon_\texttt{I}=0$, it is important to find integer
triples ($l,~m,~n$) for which $\varepsilon=0$ (see Appendix).
Secondly, we need to find the condition when all the states with non-zero values of 
$\varepsilon$ correspond to $+ve$ definite values of $\varepsilon_\texttt{I}$. For 
$K <0$, we will always have $-ve$ values of $\varepsilon_\texttt{I}$, and hence will 
never satisfy the desired condition. Thus we need to consider the case when $K>0$. 
\begin{figure}
\includegraphics[width=7cm]{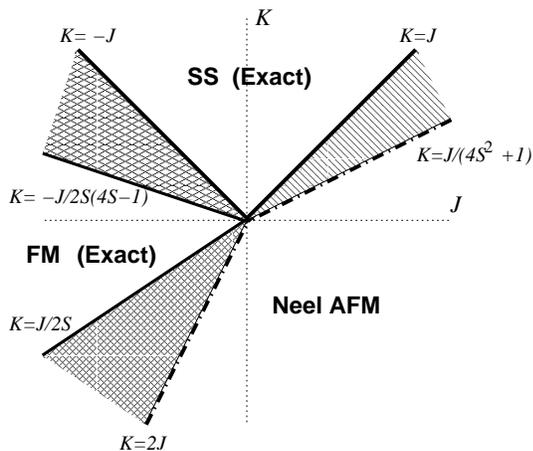}
\caption{\label{fig:phase_m1_S} The quantum phase diagram of H$_\texttt{I}$ for arbitrary 
spin-S. The bounds on the SS and the FM ground states are given by the exact sufficient 
conditions, whereas the N\'eel AFM state is given by variational bounds. The line, 
$K=-J/2$S$(4$S$-1)$, is an exact first order transition line for the FM state.}
\end{figure}

For $K>0$ and $J>0$, all $+ve$ values of $\varepsilon$ will always give $+ve$ values for 
$\varepsilon_\texttt{I}$. For all $-ve$ values of $\varepsilon$, the condition 
$-J+K|\varepsilon|>0$ should be satisfied for $\varepsilon_\texttt{I}$ to have only $+ve$ 
values. This implies $K>J/|\varepsilon|$. Since the smallest non-zero value of $|\varepsilon|$ 
is 1 ($\varepsilon=-1$ for $(l,m,n)=(1,1,1)$, and $\varepsilon=1$ for $(l,m,n)=(2,1,1)$), 
$K>J$ is the condition for the $+ve$ definiteness of $\varepsilon_\texttt{I}$ 
corresponding to all non-zero values of $\varepsilon$. Hence, for $J>0$ and $K>J$, the zero energy eigenstates 
will form the ground state of $h_\texttt{I}$. For $K>0$ and $J<0$, we need only to consider 
the states with $\varepsilon>0$. The requirement of $\varepsilon_\texttt{I} >0$ implies 
$K>|J|/\varepsilon$, which in turn implies that the condition $K>|J|$ guarantees 
$\varepsilon_\texttt{I}>0$ for all non-zero $\varepsilon$ states. And hence, the desired 
condition for $-ve$ values of $J$ is $K>|J|$. The two conditions together imply that for 
$K>|J|$, the states with $\varepsilon_\texttt{I}=0$ will form the ground state of 
$h_\texttt{I}$.

Thus $K>|J|$ is also a sufficient condition for a zero energy eigenstate of H$_\texttt{I}$ to
be the ground state. We can therefore try to construct such zero energy states by having all 
the plaquettes simultaneously in their zero energy states. As discussed above, unlike spin-1/2 
case, now there are two kinds of zero energy states for $h_\texttt{I}$ : (1) {\em trivial}, 
when S$_{13}$ (or S$_{24}$)$=0$; and (2) {\em non-trivial}, when both S$_{13}$ and S$_{24}$ 
are non-singlets. The trivial plaquette states ($i.e.$, with diagonal-singlets) will generate 
two degenerate SS states on the checkerboard, with zero eigenvalue. And for $K>|J|$, the two 
SS states will form the exact ground state of H$_\texttt{I}$.

The first non-trivial plaquette state occurs for spin-1 system, when S$_{13}=2$, S$_{24}=2$ 
and S$_{tot}=3$. The occurrence of such states is very sparse in general. See Appendix for 
further discussion, and for the related notion of perpendicularity for quantum spins. One 
can see that a non-trivial state involves all the four spins of a plaquette. That is, there 
are no {\em free} spins left (unlike a trivial state where, with one diagonal-singlet, other 
two spins can do what they want). Therefore, we can not simultaneously construct the 
non-trivial states on all corner sharing plaquettes, and thereby generate a new zero energy 
eigenstate of H$_\texttt{I}$. This leaves us with only two SS states as the zero energy 
ground state of H$_\texttt{I}$. We have not disproved the possibility of any other zero 
energy eigenstates for H$_\texttt{I}$ in a strict mathematical sense, but we believe that 
this will be the case from the arguments given above.

In the FM state where spins on each plaquette are polarized in the same way, each block 
Hamiltonian simultaneously attains maximum S$_{tot}$. Therefore, we can find the sufficient 
condition for the FM ground state by considering the situation when S$_{tot}=4$S block-state 
is the lowest energy state of $h_\texttt{I}$. We find that H$_\texttt{I}$ can have FM ground
state only for $J<0$. The sufficient condition for the existence of the FM ground state is: 
$J/2$S$<K<-J/2$S$(4$S$-1)$. We derive this by comparing the energy of S$_{tot}=4$S 
block-state with the energies of the states (S$_{tot}=4$S$-1$, S$_{13}=2$S, S$_{24}=2$S$-1$) 
for $K>0$, and (S$_{tot}=0$, S$_{13}=2$S, S$_{24}=2$S) for $K<0$ (because these block-states
compete against S$_{tot}=4$S state for the lowest energy). 

{\bf Quantum Phase Diagram :}
In Figs.~\ref{fig:phase_m1} and \ref{fig:phase_m1_S}, we show the approximate quantum phase
diagrams of spin-1/2 and spin-S cases of Model-\texttt{I}. There are three main phases in both 
cases : (1) the two-fold degenerate SS phase, (2) the FM phase, and (3) the N\'eel AFM phase. 
We have estimated the boundaries of the AFM phase by taking the N\'eel state as a variational 
choice. The variational energy (per plaquette) of Model-\texttt{I} in the N\'eel state is : 
$\varepsilon_\texttt{I}(\mbox{N\'eel}) = -4$S$^2J + 4$S$^2(4$S$^2 +1) K$. 

For $J>0$, comparing $\varepsilon_\texttt{I}(\mbox{N\'eel})$ with zero gives $K=J/(4$S$^2+1)$, 
which is the variational phase boundary between the N\'eel and the SS phases. For 
$K>J/(4$S$^2+1)$, the SS states have lower energy than the N\'eel state. However, the exact 
sufficient condition for the existence of the SS ground state is $K\geq|J|$. But it is not a 
necessary condition. This means that the SS ground state may continue even for $K<|J|$. There 
is an interesting point here. The proof for the SS ground state is based on the fact that the 
lower bound to the ground state energy of H$_\texttt{I}$ is zero for $K\geq |J|$. This does 
not imply that the zero energy states can not be the ground state for $K<|J|$. Besides, the 
SS states are always exact eigenstates of H$_\texttt{I}$. Therefore, the actual phase 
boundary, where the SS ground state gives way to another state, depends upon how the energy of 
the excited states in the SS phase vary as a function of $J$ and $K$. It is likely that the SS 
ground state extends further into the region, $K<|J|$. For $J>0$, therefore, the actual phase 
boundary for the SS ground state will lie somewhere in the shaded region (we call it a 
transition region) given by $J/(4$S$^2+1)<K<|J|$, which needs to be found numerically. We 
believe that, in the transition region, the system will pass through some intermediate complex 
states before going into the N\'eel state.

For $J<0$, the variational boundary, $K=2J$, for the N\'eel phase is obtained by comparing 
$\varepsilon_\texttt{I}(\mbox{N\'eel})$ with the FM ground state energy (the FM energy per 
plaquette: $\varepsilon_\texttt{I}(\mbox{FM})=4$S$^2J+K(4$S$^2)^2$). For $K>2J$, the FM 
state has lower energy than the N\'eel state. Again, the actual phase boundary for the FM 
ground state can be found only numerically, which is bound to lie in the region, $2J<K<J/2$S.

The transition region between the FM and the SS phases is different for spin-1/2 and spin-S
(S$\geq 1$) cases. For spin-1/2 system, as shown in Fig.~\ref{fig:phase_m1}, $K=|J|$ is the
exact phase boundary for the level-crossing transition between the SS and the FM ground states. 
This we obtain from the exact sufficient conditions on the existence of the SS and the FM
ground states which, for spin-1/2 case, leave no room for uncertainty.

For S$>1/2$, the SS phase may further extend into the region, $K<|J|$, before giving 
way to a non-singlet ground state for the $-ve$ values of $J$. The FM phase boundary, 
$K= -J/2$S$(4$S$-1)$, is however an exact first order phase boundary in the present case. 
We can show this by looking at the exact one-magnon dispersion above the FM ground state. 
The one-magnon dispersion is given as: 
\begin{eqnarray}
\varepsilon_{\rm magnon}^\pm({\bf k}) & = & \varepsilon_0 + 4K{\rm S}^2\cos{k_x}\cos{k_y}
\label{eq:magnon1} \\
& & \pm\sqrt{\varepsilon_1({\bf k})^2 + [4K{\rm S}^2\sin{k_x}\sin{k_y}]^2}\nonumber
\end{eqnarray}
where $\varepsilon_0 = -4$S$[J+K$S$(8$S$-3)]$ and $\varepsilon_1({\bf
k})=2$S$[J+4K$S$(2$S$-1)][\cos{k_x} + \cos{k_y}]$. 
The superscript $\pm$ denotes two branches of the magnon dispersion on the checkerboard
lattice, and {\bf k} takes the values within the half-Brillouin zone (same as the N\'eel AFM 
Brillouin zone on the square lattice). What one finds (for $J<0$) from Eq.~\ref{eq:magnon1} 
is that around  {\bf k}=0, $\varepsilon_{\rm magnon}^-({\bf k})\approx$ 
S$[|J|-2K$S$(4$S$-1)](k_x^2+k_y^2)/2$, which implies that for $K>|J|/2$S$(4$S$-1)$, the FM
ground state becomes unstable against magnon excitations. Across this line, for S$>1/2$, we 
expect the system to undergo a series of transitions into successive non-singlet phases 
(with decreasing total spin), and finally going into the SS phase. Along the line $K=J/2$S$ $, 
for $J<0$, the FM state is stable against the magnons. Therefore, the FM phase is likely to
extend further into the shaded region between the FM and AFM phases, before undergoing a first
order transition into the AFM state.

\section{\label{sec:model2} Discussion on Model-\texttt{II}}
{\bf Ground State of Spin-1/2 Case :}
Consider the block-Hamiltonian, $h_\texttt{II}$, whose eigen-spectrum for spin-1/2 case is 
given in Table~\ref{tab:h2}. Just as in the previous discussion, we need to find the 
conditions for the zero energy states of $h_\texttt{II}$ to become the lowest in energy so 
that H$_\texttt{II}$ can realize the SS ground state.
\begin{table}
\caption{\label{tab:h2} Eigen-spectrum of $h_\texttt{II}$ for S=1/2.}
\begin{ruledtabular}
\begin{tabular}{ccl}
 S$_{13}$ & S$_{24}$ & Energy eigenvalues\\ \hline
 0 & 0 & ~0\\
 0 & 1 & ~0\\
 1 & 0 & ~0\\ \hline
 1 & 1 & \begin{tabular}{rcl} $-2J + K$ & : & S$_{tot} = 0$\\
                            $-J + K$ & : & S$_{tot} = 1$\\
                            $J + K$ & : & S$_{tot} = 2$
         \end{tabular}
\end{tabular}
\end{ruledtabular}
\end{table}
It is clear from Table~\ref{tab:h2} that the zero energy states of $h_\texttt{II}$ become
the lowest in energy when $K>2J$ for $J>0$, and $K>-J$ for $J<0$. Since the zero energy 
states of $h_\texttt{II}$ also have at least one diagonal-singlet, just as discussed in 
the case of H$_\texttt{I}$, we can again construct two degenerate SS singlet states with 
zero energy which form the exact ground state of H$_\texttt{II}$ under the abovementioned 
conditions. For $J<0$ and $K<-J$, S$_{tot}=2$ state in (1,1) sector is the lowest energy 
state of $h_\texttt{II}$. This gives the exact FM ground state for H$_\texttt{II}$.

{\bf Ground State of Spin-S Case :}
Let us denote the eigenvalue of $h_\texttt{II}$ by $\varepsilon_\texttt{II}$. 
Let the quantum numbers S$_{tot}$, S$_{13}$, and S$_{24}$ take the values $l$, $m$, and $n$ 
respectively. Same as in the previous section, $m$ and $n$ take integer values between 0 and 
2S, and the integer $l$ is such that $|m-n| \leq l\leq m+n$. In this notation,
$\varepsilon_\texttt{II} = \frac{J}{2}\{l(l+1) - m(m+1) - n(n+1)\} + \frac{K}{4}m(m+1)
n(n+1)$. Again, when either of the diagonal bonds on a plaquette is in the singlet state 
($m$ (or $n$)$=0$), then $\varepsilon_\texttt{II} =0$. Now we try to find the condition
for which $\varepsilon_\texttt{II}\geq0$.

Since $\varepsilon_\texttt{II}$ is symmetric under the exchange of $m$ and $n$, we will
consider $n\geq m$ only. Now consider a sector of states for a given $m$ and $n$. Different
states in this sector correspond to different values of $l$. For $K\leq 0$, $h_\texttt{II}$ will 
always have some $-ve$ energy states. In order to make $\varepsilon_\texttt{II} \geq 0$, 
we need $K$ to be sufficiently positive. For $K=0$, the minimum energy state in a given 
sector corresponds to $l=n-m$ for $J>0$, and  for $J<0$, it corresponds to $l=n+m$. The 
order (according to energy) of different states in a sector remains the same for any non-zero 
value of $K$. This is because the slope of $\varepsilon_\texttt{II}$ with respect to $K$ 
depends only on $m$ and $n$, and not on $l$. Thus it is sufficient to consider the variation 
of the lowest energy state with respect to $K$, in a given sector. By making $K$ sufficiently 
$+ve$, we can raise these lowest energy states above zero. 

For $J>0$, the minimum energy for a given $m$ and $n$ is, 
$\varepsilon_\texttt{II}^{min}(m,n)= -J m(n+1) + \frac{K}{4}m(m+1)n(n+1)$. Demanding 
$\varepsilon_\texttt{II}^{min}(m,n)\geq 0$ gives us $K\geq 4J/n(m+1)$. The
lower bound to the values of $K$, for which the abovementioned condition is simultaneously
satisfied for all the sectors, is given by $n,m=1$. Thus, $\varepsilon_\texttt{II}\geq 0$ 
for $J>0$ and $K\geq 2J$. 

For $J<0$, the minimum energy in a sector is,
$\varepsilon_\texttt{II}^{min}(m,n) = J mn + \frac{K}{4}m(m+1)n(n+1)$. Demanding again that
$\varepsilon_\texttt{II}^{min}(m,n)\geq 0$ gives us $K\geq -4J/(m+1)(n+1)$. This inequality 
implies that the lower bound to $K$ again corresponds to $m,n=1$. Thus,
$\varepsilon_\texttt{II}\geq 0$ for $J<0$ and $K\geq -J$. This condition and the condition in
the previous paragraph together provide the exact conditions for the diagonal-singlet
block-states to be the lowest in energy. This further implies that two zero energy SS
eigenstates form the exact ground state of H$_\texttt{II}$.

As noticed earlier, in a given sector, the lowest energy block-state for $J<0$ corresponds 
to $l=n+m$. When $K$ is $-ve$, then $m,n=2$S is the lowest of them all. However, making 
$K$ $+ve$ would, at some point, make $l=4$S state cross above some other state. Remember 
that states belonging to same sector don't cross as a function of $K$, because they have 
same slope. Thus, for $J<0$, demanding $\varepsilon_\texttt{II}(l^*=4$S$,m^*=2$S$,n^*=2$S$)$ 
to be less than $\varepsilon_\texttt{II}(l=n+m,m,n)$ gives us the inequality: 
$K < -4J/\{m^*n^* + mn + 1 +\frac{m^*n^*(m^*+n^*) - mn(m+n)}{m^*n^* - mn}\}$. The lowest 
upper bound on $K$ for the above inequality gives the sufficient condition for $l=4$S state 
to be the lowest in energy. The lowest upper bound for $K$ is achieved when $m=2$S$-1$ and 
$n=2$S. In other words, it is a condition for level-crossing between the lowest energy state 
of $m,n=2$S sector and the same of $m=2$S$-1$, $n=2$S sector. Since $l=4$S state on a 
plaquette gives the FM state on the lattice, we have an exact sufficient condition for the 
existence of the FM ground state. That is, for $J<0$ and $K<-J/$S$(2$S$+1)$, 
H$_\texttt{II}$ has the FM ground state. 
\begin{figure}
\includegraphics[width=5.5cm]{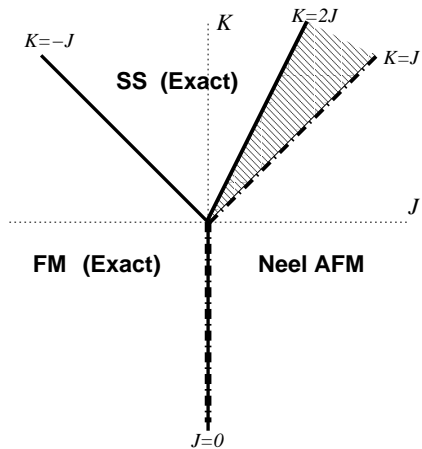}
\caption{\label{fig:phase_m2} Quantum phase diagram for spin-1/2 case of H$_\texttt{II}$. 
The line $K=-J$ is an exact level-crossing transition line between the SS and the FM states.
For $K<0$, the line $J=0$ is also an exact first order transition line.}
\end{figure}

\begin{figure}
\includegraphics[width=7cm]{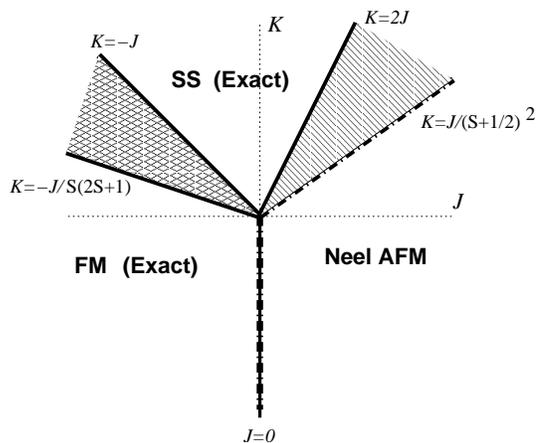}
\caption{\label{fig:phase_m2_S} The quantum phase diagram for the spin-S case of Model-\texttt{II}. 
The line $K=-J/$S$(2$S$+1)$ is an exact first order phase boundary of the FM state, and also
the line $J=0$ for $K<0$.}
\end{figure}

{\bf Quantum Phase Diagram :}
The approximate quantum phase diagrams for spin-1/2 and spin-S cases of Model-\texttt{II} are 
shown in Figs.~\ref{fig:phase_m2} and \ref{fig:phase_m2_S}. The variational energy per 
plaquette of H$_\texttt{II}$ in the N\'eel AFM state is : 
$\varepsilon_\texttt{II}(\mbox{N\'eel}) = -4$S$^2J + K$S$^2(2$S$+1)^2$. 
Comparing $\varepsilon_\texttt{II}(\mbox{N\'eel})$ with the FM ground state energy
($\varepsilon_\texttt{II}(\mbox{FM})= 4$S$^2J + K$S$^2(2$S$+1)^2$) gives
variational boundary, $J=0$, in the region, $K<0$. Incidentally, $J=0$ (for $K<0$) is also an
exact sufficient bound for the FM ground state. Therefore, $J=0$ is an exact first order phase 
boundary between the FM and the AFM ground states for $K<0$. This is unlike what we saw for
Model-\texttt{I}, where we could only give bounds on the transition region between the FM and
and AFM phases. 

For $J>0$, the variational phase boundary between the SS and the N\'eel states is given by
$K=J/($S$+1/2)^2$. The exact sufficient condition for the existence of the SS ground state, 
for $J>0$, implies that the actual line of transition below which the SS state gives way to 
some other singlet (or total S$_z=0$ state) has to lie in the region $J/($S$+1/2)^2<K<2J$, 
similar to the case of Model-\texttt{I}.

Situation for $J<0$ is again similar to that for Model-\texttt{I}. The line 
$K=-J/$S$(2$S$+1)$, is the exact first order transition line for the FM ground state. The 
one-magnon dispersion for Model-\texttt{II} is given below.
\begin{eqnarray}
\varepsilon_{\rm magnon}^\pm({\bf k}) &=& \varepsilon_2 + 
             2K{\rm S}^2(2{\rm S}+1)\cos{k_x}\cos{k_y} \label{eq:magnon2}\\
& & \pm 2{\rm S}\sqrt{\varepsilon_3({\bf k})^2+[K{\rm S}(2{\rm
S}+1)\sin{k_x}\sin{k_y}]^2}\nonumber
\end{eqnarray} 
where $\varepsilon_2 = -4J{\rm S} -2K{\rm S}^2(2{\rm S}+1)$, and $\varepsilon_3({\bf
k})=J[\cos{k_x} + \cos{k_y}]$. Again, one finds that the FM ground state becomes unstable 
against magnons for $J>0$, and for $K>-J/$S$(2$S$+1)$ when $J<0$. Thus the phase boundaries for 
the FM ground state of Model-\texttt{II} are exact first order transition lines. 

The phase boundary, $K=-J$, for spin-1/2 case is the exact level-crossing line between the FM
and the SS ground states. Similar to Model-\texttt{I}, for the case of S$>1/2$, there is a 
transition region between the FM and the SS phases, which is also expected to have a series of 
low total spin phases, finally giving way to the SS ground state. Beyond these intuitive
expectations, various shaded regions in the quantum phase diagrams need to be investigated
numerically.
\section{\label{sec:ladder} A two-leg spin-1/2 ladder model with exact AKLT ground state}
Here, we briefly present a two-leg spin-1/2 ladder model using $h_\texttt{I}$ block-Hamiltonian. 
The idea is to show an apparent relation between the AKLT construction for spin-1 chain and our 
$h_\texttt{I}$ construction for a two-leg spin-1/2 ladder. The Hamiltonian for our ladder model 
is given as : \(\mbox{H}_{\rm ladder}=\sum_{r=1}^{L}\tilde{h}_\texttt{I}(r,r+1)\), where $r$ is 
the rung index and $L$ is the total number of rungs. We assume periodic boundary condition. The 
block-Hamiltonian $\tilde{h}_\texttt{I}$ is a slightly  generalized form of $h_\texttt{I}$, as 
given below. The notation is suitably chosen for a two-leg ladder.
\begin{eqnarray}
\tilde{h}_\texttt{I}(r,r+1) &=& \frac{G}{2}\left(\S^{2}_r + \S^{2}_{r+1}\right) + 
J\S_r\cdot\S_{r+1} \nonumber\\ && + K(\S_r\cdot\S_{r+1})^2 \label{eq:h1_ladder}
\end{eqnarray}
where $\S_r = \S_{r,a} + \S_{r,b}$, is the total spin on a rung, and 
$\S_{r,a}$ and $\S_{r,b}$ are the two spins of a rung (coming from the two legs, denoted by $a$ 
and $b$ respectively). For $G=0$, the form of $\tilde{h}_\texttt{I}$ is same as $h_\texttt{I}$. 
The eigen-spectrum of $\tilde{h}_\texttt{I}(1,2)$, is given in Table~\ref{tab:h1_ladder}.
\begin{table}
\caption{\label{tab:h1_ladder} Eigen-spectrum of $\tilde{h}_\texttt{I}$ for spin-1/2.}
\begin{ruledtabular}
\begin{tabular}{ccl}
S$_1$ & S$_2$ & Energy eigenvalues\\ \hline
0 & 0 & ~0\\
0 & 1 & ~G\\
1 & 0 & ~G\\ \hline
1 & 1 & ~\begin{tabular}{lcl}
         $2G - 2J + 4K$ & : & S$_{tot} =0$\\
         $2G -J + K$ & : & S$_{tot} =1$\\
         $2G +J + K$ & : & S$_{tot} =2$
        \end{tabular}
\end{tabular}
\end{ruledtabular}
\end{table}
We can construct a large number of exact eigenstates for this model, but we consider 
the ground state first. 

For $G>0$, the block-states (S$_1$,S$_2)=(0,1)$ or (1,0) will always cost energy. Thus the 
block-states that compete for the lowest energy, come from (0,0) and (1,1) sector. The
state (0,0) will give rise to the rung singlet (RS) state for H$_{\rm ladder}$, which is a 
rather commonly occurring ground state in various two-leg ladder models. The sufficient 
conditions for the RS ground state are : (1) ($J<-G$ and $K>-J-2G$), (2) ($-G<J<3G$ and 
$K>(J-G)/2$), and (3) ($J>3G$ and $K>J-2G$). These conditions ensure that the energies of all 
the block-states in (1,1) sector are greater than zero.  The sufficient condition for the FM 
ground state is : $J<K<-J-2G$. The other well-known ground state which can be exactly realized 
for this model is the AKLT singlet state~\cite{AKLT_Lett}. This will be described in the 
following paragraph. The RS and the AKLT states can be written as:
\begin{eqnarray}
|\textrm{RS}\rangle &=& \otimes\prod_{r=1}^L |s,r\rangle\label{eq:RS}\\
|\textrm{AKLT}\rangle &=& Tr\prod_{r=1}^L|\Psi,r\rangle \label{eq:AKLT}
\end{eqnarray}
where $|s,r\rangle = (\uparrow^a_r\downarrow^b_r - \downarrow^a_r\uparrow^b_r)/\sqrt{2}$, and
$|\Psi,r\rangle$ is a matrix wavefunction~\cite{Mikeska1} in the triplet sector of the total 
spin on $r^{th}$ rung. It can be written as:\[
|\Psi,r\rangle = \left(\begin{array}{cc} |0,r\rangle & -\sqrt{2}|\overline{1},r\rangle\\ 
                                     \sqrt{2}|1,r\rangle & -|0,r\rangle
                   \end{array}\right)
\]
Here $|1,r\rangle$, $|0,r\rangle$ and $\overline{1},r\rangle$ are the spin-1 states on a rung.
In Eq.~\ref{eq:AKLT}, $Tr$ denotes the trace of the matrix-product, which gives the AKLT state
for a closed chain. Without trace, the matrix-product will give the AKLT state for an open 
spin-1 chain (equivalently for the ladder in the present case). 

For $J=3K$, S$_{tot}=0$ and 1 states in (1,1) sector are degenerate while S$_{tot}=2$ has
different energy. Thus for $J=3K$, $\tilde{h}_\texttt{I}$ actually acts like the AKLT
projection operator in (1,1) sector. Therefore, when $G>0$, the model has the following 
three ground states corresponding to three different segments on $J=3K$ line. It has the FM 
ground state for $K<-G/2$, the RS ground state for $-G/2<K<G$, and the AKLT ground state for 
$K>G$. The point $K=-G/2$ is the level-crossing point between the FM and the RS ground states, 
and $K=G$ is the same for the RS and the AKLT ground states. At both these points, the ground 
state is two-fold degenerate, and a domain wall is an exact localized excitation of energy $G$. 
This is because the block-state on the plaquette at the domain wall (whose one rung belongs to 
the RS state and the other to the AKLT or FM states), is either (0,1) or (1,0).

For $G<0$, the block-states, (0,1) and (1,0), are $-ve$ energy states. Their competition with
the states in (1,1) sector gives the sufficient conditions for various exact ground states. For
$K-|G|>|J|$, \{(0,1),(1,0)\} states are lowest in the energy, and the exact ground state of
H$_{\rm ladder}$ consists of configurations with singlets on alternate rungs, and the same 
for triplets. Let us call it as the Alternating-Singlet-Triplet (AST) state. It can be written 
explicitly as:
\begin{eqnarray}
|{\rm AST}_1\rangle &=& 
 |s,1\rangle\otimes|t,2\rangle\otimes|s,3\rangle\otimes|t,4\rangle\otimes\cdots \label{eq:AST1}\\
|{\rm AST}_2\rangle &=& 
 |t,1\rangle\otimes|s,2\rangle\otimes|t,3\rangle\otimes|s,4\rangle\otimes\cdots \label{eq:AST2}
\end{eqnarray}
where $|s,r\rangle$ and $|t,r\rangle$ denote the singlet and a triplet state on $r^{th}$ rung. Since 
there are $2\times(L/2)^3$ degenerate AST states, this ground state has an extensive entropy.
The sufficient condition for the FM ground state, when $G<0$, is : $J<0$ and $J<K<-J+|G|$.
Again, for $J=3K$, the AKLT state is an exact ground state of H$_\textrm{ladder}$ when $J>0$,
and for $J<0$, the FM state is the exact ground state. This is still more closer to the spin-1 
AKLT model, because depending upon the sign of $J$, the ground state is either FM or AKLT,
exactly like the AKLT chain. In fact in the limit of $G\rightarrow -\infty$, H$_\textrm{ladder}$ 
gets completely projected onto the spin-1 AKLT model.

Finally, about a large number of other exact eigenstate of H$_\textrm{ladder}$. Start with the
RS state as the reference eigenstate. Changing the singlet on a rung into a triplet gives a new 
eigenstate with energy $2G$. We may call these states as one-triplet states. These are localized 
eigenstates. Next, we can have two rungs having triplets. This case has two different situations 
depending upon whether the triplet-rungs are nearest neighbors or not. If they are not, then it's 
like two independent one-triplet states. If yes then, two neighboring triplet-rungs will form 
plaquette eigenstates of (1,1) sector. Thus we have two types of eigenstates in two-triplet 
sector (with reference to the RS state). This is similar to, say, two independent quasi-particles 
and to their bound-states in a given system. For a state with three rungs having triplets, we 
have three one-triplet states, two `one-triplet + a bound state' states, and third possibility
of a three site open boundary spin-1 bilinear-biquadratic problem (which becomes AKLT only for 
$J=3K$).

In simple terms, H$_\textrm{ladder}$ is a one dimensional model, with site variables which are 
either spin-0 or spin-1, with different local energies (0 or $2G$, respectively). If the site
variable is spin-0, then it is {\em inert}, because the Hamiltonian does nothing to it. If it
is spin-1, then it can interact with neighboring site variables, and hence an {\em active}
variable. Therefore, for a given number of active sites, the problem can be resolved into 
different sectors of a set of independent open boundary spin-1 problems separated by 
inert sites. For example, the N-active site problem can be resolved into N one-active site 
problems, or `(N-1) one-active site problems + 1 two-active site problem', and so on, finally 
into an N-active site problem. For spins greater than 1/2, singlet-rungs will still be inert,
but now the number of active rungs (or sites) becomes larger. One can still find the exact 
conditions for the FM and the RS ground states, but the problem otherwise becomes more involved. 
\section{\label{sec:conclusion} conclusion}
We, now, conclude this work with a summary, and with some general remarks. We have presented
two exactly solvable quantum spin models, Models-\texttt{I} and \texttt{II}, with 
bilinear-biquadratic exchange interactions on checkerboard lattice, for arbitrary spin-S. We 
have shown that two degenerate SS singlet states form an exact ground state of these models. 
We have also found the sufficient conditions for the existence of both the SS and the FM 
ground states, and presented approximate quantum phase diagrams. Inspired by our discussion 
on Model-\texttt{I}, we have also introduced the notion of perpendicularity for quantum 
spin vectors, which is discussed in the Appendix.

We have, further, discussed a simple two-leg spin-1/2 ladder model, and pointed out its 
apparent connection with a general bilinear-biquadratic spin-1 chain. Particularly 
interesting, for a certain ratio of the exchange couplings, is its similarity with the 
AKLT spin-1 chain, for which it can realize the exact AKLT singlet ground state (which shows 
level-crossing transition with the FM or the RS ground states). It can also have highly 
entropic, exact ground state of AST type (see Eqs.~\ref{eq:AST1} $\&$~\ref{eq:AST2}).

It is important to note that the block-construction scheme of Models-\texttt{I} and 
\texttt{II}, is directly applicable to the models of corner-sharing octahedra, of which 
the pyrochlore lattice could be an interesting case. A plausible case of further study, for 
Models-\texttt{I} and \texttt{II}, with a realistic interest, is the coupling to phonons. 
It is possible that a suitable spin-phonon interaction will favor one of the two SS states by 
inducing antiferro-distortive arrangement of plaquettes on the checkerboard, though there may 
still be two such degenerate states. Therefore, the requirement of large $K/J$ for the SS 
ground state may not be a limitation, in a realistic situation, if at all. This, in a way,
would be similar to the relation between SrCu$_2$(BO$_3$)$_2$ and the SS model.

A brief remark about the classical spins. In this case (with an appropriate
rescaling of $K\rightarrow K/$S$^2$), there is no SS phase. Instead, the zero energy
sector has a huge classical degeneracy (which is typical of a frustrated classical
antiferromagnet). This degeneracy arises due to the fact that each dimer (in an SS like
arrangement) requires to have two oppositely oriented spins, independent of the spins in all 
other dimers. The shaded area between the SS and the N\'eel state (in
Figs.~\ref{fig:phase_m1_S} and \ref{fig:phase_m2_S}) will become the region of zero energy 
states in the classical limit.

Finally, it is interesting to mention that the exact ground states of the models, that are presented 
here, are superstable eigenstates. The superstability of an eigenstate of an operator signifies 
the stability of its being eigenstate against the inclusion of extra interactions which do not 
commute with the original operator~\cite{super_SS}. The dimer-singlet states, of which the SS
states are one example, manifest this property rather well~\cite{bkumar}. There is also a further 
stability against the {\em exchange disorder}. For example, the existence of the SS ground state 
for Model-\texttt{I} requires one to have $K>|J|$ on each plaquette. However, $J$ and $K$ on 
different plaquettes can still be different (say, $J$ having random sign), as long as each 
plaquette fulfills the sufficient condition. This is because the sufficient condition is a local 
(plaquette) condition. By the same argument, the FM ground state is also stable against the 
disorder in $J$ and $K$, as long as the sufficient conditions are respected. We remark that
such a stability against {\em constrained} disorder is an important physical aspect of the 
superstability property.
\begin{acknowledgments}
The author sincerely thanks Frederic Mila and Valeri Kotov for useful comments 
and encouragement. He also likes to acknowledge the financial support from the Swiss National 
Funds.
\end{acknowledgments}
\appendix*
\section{\label{append:perp} Perpendicular States for quantum spins}
The discussion on Model-\texttt{I} led us to the following question. For a system of two
quantum spins, $\S_A$ and $\S_B$, what are the states for which $\S_A\cdot\S_B$ is zero? 
To be precise, it's about finding an eigenstate $|\psi\rangle$ of $\S_A\cdot\S_B$ such 
that $\S_A\cdot\S_B|\psi\rangle =0$. This naturally leads us to formulate the notion of 
{\em perpendicularity of quantum spin vectors}. We develop and elaborate upon this concept in
the following.

Classically, the scalar product of two perpendicular vectors is zero. For quantum spins 
vectors, this is not an obvious proposition. This is because the states describing a quantum 
spin vector carry information only about the total length and the length of one of its 
components (say, $z$-component). Since the
uncertainty about the other two components of a spin vector is inherent in the quantum
description, there is no strict sense of {\em direction} for a quantum spin vector. The
simplest demonstration of this fact can be in seen in the relation : $\S\times\S = i\S$,
where the right hand side is a direct consequence of the quantum mechanical uncertainty in the
transverse components ($\hbar$ is taken to be 1). 
And hence, the notion of perpendicularity of two quantum spin vectors
is not something straightforward. In fact for the extreme quantum case of spin-1/2, the
eigenvalues of the scalar product of two spins are 1/4 (triplet) or $-3/4$ (singlet), both 
non-zero. The question now is whether it is possible to find a zero eigenvalue state of the
scalar product operator of two spins. The answer to this question lies in the integer (or
half-integer) solutions of an equation that we may like to call as the quantum analog of 
the Pythagorean relation.

Let $l$, $m$ and $n$ be the spin quantum numbers for operators $\S_{A+B}$ ($=\S_A + \S_B$),
$\S_A$ and $\S_B$ respectively. Since $\S_A\cdot\S_B = \frac{1}{2}[\S_{A+B}^2 - \S_A^2 - 
\S_B^2]$, the eigenvalue of a state specified by $l$, $m$ and $n$ is
$\varepsilon(l,m,n)=\frac{1}{2}[l(l+1) - m(m+1) - n(n+1)]$. Our interest, here, is to find an 
eigenstate with $\varepsilon(l,m,n)=0$. We call such an eigenstate as the {\em perpendicular state}, 
and denote it by $|\perp\rangle$ (to symbolize its equivalence to the perpendicularity of classical 
vectors). This requires us to find the integer or half-integer solutions (because spin quantum 
numbers can be integer or half-integer) to the equation:
\begin{equation}
l(l+1) = m(m+1) + n(n+1)
\label{eq:q_pytha}
\end{equation}
We call Eq.~\ref{eq:q_pytha} as the quantum analog of the Pythagorean equation: 
$l^2 = m^2 + n^2$. Taking $l$, $m$, $n$ to be $L/2$, $M/2$ and $N/2$ respectively 
(where $L$, $M$, $N$ are $+ve$ integers), Eq.~\ref{eq:q_pytha} can be re-written as: 
$L(L+2) = M(M+2) + N(N+2)$, and now we need to find only the integer solutions for 
$L$, $M$ and $N$. In this notation, the odd integers correspond to the half-integer spins and 
the even integers to the integer spins.

One interesting thing that we can say for sure is that, for odd integer values of $M$ and $N$, 
the above equation will never be satisfied for any integer value of $L$. 
This means that {\em two half-integer spins can never be in a perpendicular state}! 
This is an exact proposition. And for one more time, it puts half-integer spins in a special 
place. However, the same is not true for other three cases, that is, ($M$,$N$) $=$ (odd,even),
(even,odd) and (even,even). 

The first integer solution to the equation is $(L,M,N)=(7,3,6)$,
which corresponds to the spin quantum numbers, $(l,m,n)=(7/2, 3/2, 3)$. By first we mean the 
smallest value of $m$ that satisfies Eq.~\ref{eq:q_pytha} for some suitable values of $n$ and 
$l$. Note that for a given solution $(l,m,n)$ of Eq.~\ref{eq:q_pytha}, $(l,n,m)$ is also a 
solution, because Eq.~\ref{eq:q_pytha} is symmetric under the exchange of $m$ and $n$. We will, 
therefore, consider solutions with $n\geq m$ only.  This is just a convention for enumerating 
various perpendicular states. Thus, the first {\em perpendicular} state is 
$|$S$_{A+B}=7/2,~$S$_A=3/2,~$S$_B=3\rangle$. Then next few perpendicular states are : 
$(l,m,n)=$(3, 2, 2),~(17/2, 5/2, 8),~(6, 3, 5),~(11/2, 7/2, 4),~\dots. 
And, of course, there are many more. One can enumerate successive states by a one-line 
Mathematica program.

A perpendicular state, $|\perp\rangle$, also satisfies many relations that hold true for 
classical perpendicularity, which otherwise will not be satisfied by an arbitrary quantum spin 
state. This happens to be true precisely because $\S_A\cdot\S_B|\perp\rangle =0$. We describe some 
of these identities below. First we consider the cross-product, $\S_A\times\S_B$, of two
quantum spins. One can show the following identity for the cross-product operator.
\begin{equation}
(\S_A\times\S_B)^2 = \S_A^2\S_B^2 - (\S_A\cdot\S_B)^2 -\S_A\cdot\S_B
\label{eq:q_indentity1}
\end{equation}
Interestingly, in the perpendicular state, 
$(\S_A\times\S_B)^2|\perp\rangle = \S_A^2\S_B^2|\perp\rangle$, which is 
same as for two perpendicular classical vectors. Of course, here $\S_A^2 =$S$_A$(S$_A$ +1), and
similarly for $\S_B^2$. In the following, we describe some more relations that are satisfied by
$|\perp\rangle$, in the spirit of classical perpendicularity.

For classical vectors, the identities, $\S_A\cdot(\S_A\times\S_B)=0$ 
and $\S_B\cdot(\S_A\times\S_B)=0$ are always true, regardless of the fact whether $\S_A$ is
perpendicular to $\S_B$ or not. We will show that, for quantum spin vectors, these are not true
in general. However,  in the state $|\perp\rangle$, it happens to be true. In other words, the
cross-product operator of two quantum spins can be perpendicular to the spins themselves only 
in the perpendicular state of the two spins. The proof is as follows. For quantum spin vectors, 
one can show that the following identity holds true.
\begin{equation}
\S_A\cdot(\S_A\times\S_B) = -(\S_A\times\S_B)\cdot\S_A = i\S_A\cdot\S_B
\label{eq:q_identity2}
\end{equation}
There are two interesting facts about Eq.~\ref{eq:q_identity2}. One is that the scalar product of
$\S_A$ and $\S_A\times\S_B$ is anti-commuting. We may like to call it as {\em scalar product 
anti-commutation}. And secondly, $\S_A\cdot(\S_A\times\S_B)$ is non-zero, but proportional to 
$\S_A\cdot\S_B$. For the state $|\perp\rangle$, this second fact implies that 
$\S_A\cdot(\S_A\times\S_B)|\perp\rangle=0$, and similarly for $\S_B\cdot(\S_A\times\S_B)$.

We also find that $\S_A\cdot(\S_A\times(\S_A\times\S_B))|\perp\rangle=0$ and
$\S_B\cdot(\S_A\times(\S_A\times\S_B))|\perp\rangle=-\S_A^2\S_B^2|\perp\rangle$.
This is as it should be in the classical sense. The above results 
follow from the identities: 
\begin{eqnarray}
\S_A\cdot(\S_A\times(\S_A\times\S_B)) & = & -\S_A\cdot\S_B,\label{eq:q_identity3_1}\\ 
\S_B\cdot(\S_A\times(\S_A\times\S_B)) & = & -(\S_A\times\S_B)^2 \label{eq:q_identity3_2} 
\end{eqnarray}
It is encouraging to see such close correspondence between the classical perpendicularity
and the quantum state, $|\perp\rangle$. Though one may still like to further understand the
physical import of a perpendicular state.
\bibliography{manuscript}
\end{document}